\documentclass[times,twocolumn,final]{elsarticle}
\usepackage{framed,multirow}

\usepackage{amssymb}
\usepackage{latexsym}

\usepackage{url}
\usepackage{xcolor}

\usepackage{hyperref}

\definecolor{newcolor}{rgb}{.8,.349,.1}

\usepackage[utf8]{inputenc}
\usepackage[T1]{fontenc}
\usepackage{bm}
\usepackage{multirow}
\usepackage{eucal}
\usepackage{enumitem}
\usepackage[stable,bottom,hang]{footmisc}
\usepackage{color}
\usepackage{hyperref}
\usepackage{amssymb}
\usepackage{amsmath}
\usepackage{algorithm}
\usepackage{algorithmic}
\usepackage{graphicx}
\usepackage{ragged2e}
\usepackage{algorithm}
\usepackage{algorithmic}
\usepackage[numbers]{natbib}

\usepackage{tabularx,booktabs}
\newcolumntype{C}{>{\centering\arraybackslash}X} 
\begin{document}
\begin{frontmatter}

\title{Explainable-by-design Semi-Supervised Representation Learning for COVID-19 Diagnosis from CT Imaging}
\author[1,2,16]{Abel D\'{i}az Berenguer}
\author[1,2,3]{Hichem Sahli}
\author[1,2,3]{Boris Joukovsky}

\author[1,2,3]{Maryna Kvasnytsia}
\author[1,2,3]{Ine Dirks}
\author[1,2]{Mitchel Alioscha-Perez}
\author[1,2,3]{Nikos Deligiannis}

\author[1,2,3]{Panagiotis Gonidakis}
\author[1,2,3]{Sebastián Amador Sánchez}
\author[1,2,3]{Redona Brahimetaj}
\author[1,2,3]{Evgenia Papavasileiou}
\author[1,2]{Jonathan Cheung-Wai Chan}
\author[1,2, 17]{Fei Li}
\author[1,2, 18]{Shangzhen Song}
\author[1,2, 18]{Yixin Yang}

\author[1,4,15]{Sofie Tilborghs}
\author[1,4,15]{Siri Willems}
\author[1,4,15]{Tom Eelbode}
\author[1,4,15]{Jeroen Bertels}
\author[1,4,15]{Dirk Vandermeulen}
\author[1,4,15]{Frederik Maes}	
\author[1,4,15]{Paul Suetens}

\author[1,5]{Lucas Fidon}
\author[1,5]{Tom Vercauteren}

\author[1,4,6,15]{David Robben}
\author[1,6]{Arne Brys}
\author[1,6]{Dirk Smeets}

\author[1,7]{Bart Ilsen}
\author[1,7]{Nico Buls}
\author[1,7]{Nina Watt\'e}
\author[1,7]{Johan de Mey}

\author[1,8,9]{Annemiek Snoeckx}
\author[1,8,10,11]{Paul M. Parizel}

\author[1,12]{Julien Guiot}
\author[1,13]{Louis Deprez}
\author[1,13]{Paul Meunier}

\author[1,14]{Stefaan Gryspeerdt}
\author[1,14]{Kristof De Smet}

\author[1,2,3]{Bart Jansen}
\author[1,2,3]{Jef Vandemeulebroucke}

\address[1]{icovid consortium}
\address[2]{Vrije Universiteit Brussel (VUB), Department of Electronics and Informatics (ETRO), Brussels, Belgium}
\address[3]{Interuniversity Microelectronics Centre (IMEC), Heverlee, Belgium}
\address[4]{KU Leuven, Department of Electrical Engineering, ESAT/PSI, Leuven, Belgium}
\address[5]{School of Biomedical Engineering \& Imaging Sciences, King's College London, London, United Kingdom}
\address[6]{icometrix, Leuven, Belgium}
\address[7]{Vrije Universiteit Brussel (VUB), Universitair Ziekenhuis Brussel (UZ Brussel), Brussels, Belgium}
\address[8]{University of Antwerp (UA), Antwerp, Belgium}
\address[9]{Antwerp University Hospital (UZA), Antwerp, Belgium}
\address[10]{Royal Perth Hospital, Perth, Australia}
\address[11]{University of Western Australia, Perth, Australia}
\address[12]{University Hospital of Liège, Department of Pneumology, Liège, Belgium}
\address[13]{University Hospital of Liège, Department of Radiology, Liège, Belgium}
\address[14]{AZ Delta, Department of Radiology, Roeselare, Belgium}
\address[15]{UZ Leuven, Medical Imaging Research Center, Leuven, Belgium}
\address[16]{Universidad de las Ciencias Informáticas (UCI), La Habana, Cuba}
\address[17]{Northwestern Polytechnical University, Xi'an, China}
\address[18]{XiDian University, Xi'an, China}
   
\begin{abstract}
Our motivating application is a real-world problem: COVID-19 classification from CT imaging, for which we present a explainable Deep Learning approach based on a semi-supervised classification pipeline that employs variational autoencoders to extract efficient feature embedding. We have optimized the architecture of two different networks for CT images: (i) a novel conditional variational autoencoder (CVAE) with a specific architecture that integrates the class labels inside the encoder layers and uses side information with shared attention layers for the encoder, which make the most of the contextual clues for representation learning, and (ii) a downstream convolutional neural network for supervised classification using the encoder structure of the CVAE. With the explainable classification results, the proposed diagnosis system is very effective for COVID-19 classification. Based on the promising results obtained qualitatively and quantitatively, we envisage a wide deployment of our developed technique in large-scale clinical studies. Code is available at \url{https://git.etrovub.be/AVSP/ct-based-covid-19-diagnostic-tool.git}
\end{abstract}

\end{frontmatter}

\section{Introduction}

In December $2019$, an outbreak of pneumonia of unknown causes was first reported in the Hubei province, China. It was soon discovered to be associated with a novel coronavirus called severe acute respiratory syndrome coronavirus 2 (SARS-CoV-2). The virus and associated coronavirus disease $2019$ (COVID-19) spread globally and was declared a pandemic, infecting nearly $7$ million people on $7$ June $2020$.

The principal approaches to control the spread of the virus are early detection and isolation. Currently, real-time reverse transcription polymerase chain reaction (RT-PCR) assay from pharyngeal swabs is considered the reference diagnostic test for COVID-19. However, its limited availability and  lengthy turnaround time have led many to explore complementary methods as a frontline diagnostic tool.

CT imaging has been widely used during the COVID-19 pandemic. CT findings associated with COVID-19 include bilateral pulmonary parenchymal ground-glass and consolidative pulmonary opacities, sometimes with a rounded morphology and a peripheral lung distribution~\citep{Bernheimetal2020}. Mild or moderate progression of the disease is manifested by increasing extent and density of lung opacities~\citep{Chungetal2020}.

CT imaging is recommended for rating the disease severity and triage of patients for referral to RT-PCR testing in case of shortage of testing material. Its use for diagnosis in patients suspected of having COVID-19 and presenting with mild clinical features is not indicated unless they are at risk for disease progression~\citep{Rubinetal2020}. A recent meta-analysis~\citep{Xuetal2020}, covering $16$ independent studies, reported pooled sensitivity of $92\%$ (range $61\%$ to $99\%$) with only two studies reporting specificity, $25\%$ and $33\%$, respectively.

Several systems standardizing the assessment and reporting of COVID-19 on non-enhanced chest CT have been proposed~\citep{Prokopetal2020, Simpsonetal2020, Salehietal2020}. An example is the COVID-19 Reporting and Data System (CO-RADS)~\citep{Prokopetal2020}, assessing the suspicion for pulmonary involvement of COVID-19 on a scale from 1 (very low) to 5 (very high). In a study investigating the use of CO-RADS, an average area under the curve (AUC) of 0.95 ( CI 0.91-0.99) for predicting clinical diagnosis was reported, along with an interobserver variation in terms of the Fleiss' kappa score of 0.47~\citep{Prokopetal2020}.

The aim of the current work is to develop a method for  automated diagnosis of COVID-19 from CT imaging capable of improving the reproducibility and diagnostic accuracy obtained through manual reading. We present a complete semi-supervised classification pipeline, that that is explainable by design. We have optimized the architecture of two different networks for CT images. A generative model based on conditional variational autoencoders (CVAE)~\citep{Sohnetal2015},~\citep{Ivanovetal2019}, relying on two inputs: the CT images and the class labels along with a representation learning algorithm which incorporates side information in the form of an additional image modality at training time, representing the segmented lung lesions, to produce a more informed representation learning.

Once we learn the CVAE encoder that maps images into a disentangled latent space, the weights of the encoder are used to initialize a downstream convolutional neural network (CNN) for supervised classification. Feature aggregation layers are added on top of the  CVAE's encoder base, and the resulting network is trained, in conjunction with a focal loss function, to classify CT scans. While our method is generic and hence widely applicable, we apply the pipeline to the problem of classifying CT volumes into COVID-19 positive versus negative patients. With the explainable classification results, our system largely simplifies and accelerates the diagnosis process for manual reading.

The remainder of this paper is organized as follows. In Section~\ref{sec:Related Work} we give an overview of the related work. In Section\ref{sec:Data} we present the collected COVID-CT data-set. The proposed diagnosis system is detailed in Section~\ref{sec:Diagnosis}. The experimental results are addressed in Section~\ref{sec:Experiments}. Finally, conclusions are drawn in Section~\ref{sec:conclusions}.

\section{Related work}
\label{sec:Related Work}
Several recent works targeted the development of deep learning models to classify CT images and predict whether the CTs are positive for COVID-19. The recent state-of-art models can be categorized into (a)~direct CT classification, (b)~COVID-19 lesion detection followed by classification, or (c)~CT classification followed by COVID-19 lesion segmentation.


For the purpose of CT scan classification into COVID-19 positive versus non-infected patients, the authors of~\citep{Singhetal2020} proposed a convolutional neural network (CNN) consisting of $3$ convolutional layers, each followed by ReLU and max-pooling, and three fully-connected layers followed by SoftMax for classification. The CNN hyperparameters have been tuned using a multi-objective differential evolution algorithm~\citep{ZhangSanderson2009} considering a fitness function combining sensitivity and specificity.

Hasan {\em et al.}~\citep{Hasanetal2020} proposed the combination of CNN features with hand-crafted texture features (Q-deformed entropy), extracted from individual CT scans, along with a feature selection process and a long short-term memory (LSTM) neural network classifier for discriminating between COVID-19, pneumonia and healthy CT lung scans. The CNN features have been obtained with an architecture consisting of $4$ convolutional layers followed by max-pooling and a fully-connected last layer.

The authors of~\citep{HeXetal2020} proposed a self-supervised transfer learning approach where contrastive self-supervised learning~\citep{Chenetal2020} is integrated into a transfer learning process to adjust the network weights pre-trained on a source data-set. The authors evaluated the efficacy of different transfer learning strategies with different source data-sets (ImageNet and the Lung Nodule Malignancy (LNM) data-set~\citep{Shanetal2020}) and network architectures such as ResNet~\citep{HeKetal2016}, DenseNet~\citep{Huangetal2017} and proposed an architecture for binary classification (CRNet) consisting of a combination of a 2D-convolution with a ReLU activation repeating four times. For the purpose of interpretability of the model, the authors adopted the Gradient-weighted Class Activation Mapping (Grad-CAM)~\citep{Selvarajuetal2017} method to visualize the important regions leading to the decision of the proposed model. They evaluated their approach on a publicly available data-set~\citep{Zhaoetal2020}.

Butt {\em et al.}~\citep{Buttetal2020} considered the classification of CT-scans in three classes: COVID-19, influenza viral pneumonia, or no-infection. They proposed a ResNet-18~\citep{HeKetal2016} network structure for image feature extraction. The output of the convolution layer was flattened to a $256$-dimensional feature vector and then converted into a $16$-dimensional feature vector using a fully-connected layer, to which they concatenated a location-attention vector, estimated as the relative distance-from-edge. Next, three fully-connected layers were followed to output the final classification result together with the confidence score. Image patches from the same cube (center image, together with the two neighbors) voted for the type and confidence score of the candidate, and the overall analysis report for one CT sample was calculated using the Noisy-or Bayesian function~\citep{Oniskoetal2001}.

Song {\em et al.}~\citep{Yingetal2020} used a pre-trained ResNet-50~\citep{HeKetal2016} architecture, on which the Feature Pyramid Network (FPN)~\citep{Linetal2017a} has been coupled to the recurrent attention convolutional neural network~\citep{Fuetal2017} to learn discriminative region attention. The  results were aggregated using mean pooling for final classification. 

In~\citep{LiLetal2020}, the authors  proposed COVNet, a 3D deep learning framework for the classification of CT scans as COVID-19, community acquired pneumonia (CAP), and non-pneumonia. COVNet consists of a RestNet50~\citep{HeKetal2016} as backbone. It takes a series of CT slices as input and generates features for the corresponding slices.  The extracted features from all slices are then combined by a max-pooling operation.  The final feature map is fed to a fully-connected layer and SoftMax to generate a probability score for each class. The authors adopted the Gradient-weighted Class Activation Mapping (Grad-CAM)~\citep{Selvarajuetal2017} method to visualize the important regions leading to the decision of the COVNet model.

The authors of~\citep{Huelal2020} proposed a multi-scale learning scheme to cope with variations of the size
and location of COVID-19 lesions.  They used a CNN architecture with $5$ convolutional layers, and
 applied a spatial aggregation with a Global Max Pooling (GMP) operation to obtain categorical scores. 
 To deal with the problem of imbalanced classification, they added a class-balanced weighting factor to the cross-entropy loss. They also extracted the position of the lesions or inflammations caused by the COVID-19 using the integrated gradients feature attribution method~\citep{Sundararajanetal2017} that assigns to the pixels an importance score representing how much the pixel value adds or subtracts from the network output.

Wang {\em et al.}~\citep{Wangetal2020} proposed DeCoVNet, a 3D deep CNN for COVID-19-positive and COVID-19-negative classifications. DeCoVNet consists of three stages and takes as input the CT volume and its 3D lung mask. The first stage is a vanilla 3D convolution with a kernel size of $5 \times 7 \times 7$, a batchnorm layer and a pooling layer. The second stage is composed of two 3D residual blocks (ResBlocks). The third stage is a progressive classifier composed of three 3D convolution layers and a fully-connected (FC) layer with the SoftMax activation function. The authors also proposed a weakly-supervised lesion localization architecture, by selecting overlapping regions inferred from the DecCovNet after applying the class activation mapping (CAM)~\citep{Zhouetal2016} and regions obtained using a 3D connected component method~\citep{Liaoetal2019} to the CT scans.

For the purpose of automated COVID-19 lesions detection and classification, Gozes {\em et al.}~\citep{Gozesetal2020} proposed a system that analyzes CT scans at two levels: a first level that analysis the 3D lung volume, using a commercial off-the-shelf software\footnote{RADLogics Inc. \url{www.radlogics.com/}}, and detects nodules and small opacities, and a second level that analyzes each 2D slice separately, using a pre-trained Resnet-50~\citep{HeKetal2016} for detecting COVID-19 related abnormalities, fine-tuned with annotated per slice normal and abnormal (COVID-19) data. To overcome the limited amount of cases, the authors employed data augmentation techniques. To mark a case as COVID-19 positive, the authors calculated the ratio of positive detected slices out of the total slices of the lung (positive ratio). A positive case-decision is made if the positive ratio exceeds a pre-defined threshold. In a follow-up abnormality localization step, given a new slice classified as positive, the authors extracted network-activation maps using the Grad-CAM technique~\citep{Selvarajuetal2017} to produce a visual explanation for the network decision.

Following the same idea of lesion segmentation followed by lesions classification, Shi {\em et al.}~\citep{Shietal2020} proposed a system that first preprocesses the CT volume to obtain the segmentations of both lesions and lung using the VB-Net network of~\citep{Shanetal2020}. Then an infection Size Aware Random Forest method (iSARF) is applied, in which subjects were first categorized into groups with different ranges of infected lesion size. In each group random forests is adopted for final classification using a series of handcrafted features, including infected lesion number, volume, histogram distribution, and surface area, estimated from the lesions and lung.

Jin {\em et al.}~\citep{Jinetal2020} also proposed a combined lesion segmentation step, via 3D U-net++~\citep{Zhouetal2018}, followed by a lesion classification step using ResNet-50~\citep{HeKetal2016}. The two models have been trained separately. The classification model takes as input dual-channel information, i.e., the lesion regions and their corresponding segmentation masks (obtained from the previous segmentation model) for the final classification results (positive or negative).


Considering a classification followed by a lesion segmentation system, Wu {\em et al.}~\citep{Wuetal2020} proposed Res2Net~\citep{Gaoetal2019} as classifier trained using the image mixing approach~\citep{Zhangetal2018}. In a second step, the authors applied an encoder-decoder based segmentation model for detecting the lesion areas. The encoder is based on the VGG-16~\citep{SimonyanZisserman2014} backbone, without the last two fully-connected layers, to which the authors added two sequential Grouped Atrous Modules and a max pooling function between them. The decoder has $5$ side-outputs, with an Attentive Feature Fusion strategy to aggregate the feature maps from different stages and predict the side-output of each stage~\citep{Wuetal2020}. For the purpose of interpretability, the authors adopted the Activation Mapping~\citep{Selvarajuetal2017} method to visualize the important regions leading to the decision of the classification model.

A multi-task deep learning model has been proposed by Amyar {\em et al.}~\citep{Amyaretal2020}. The model allows, (i)~a COVID-19 vs non-COVID-19 classification, (ii)~COVID-19 lesion segmentation, and (iii)~image reconstruction. The architecture consists in a common encoder for the three tasks which takes a CT scan as input, and its output is then used to the reconstruction of the image via a first decoder, to the segmentation via a second decoder, and to the classification via a multi-layer perceptron. The authors made use of a 2D U-net~\citep{Ronnebergeretal2015} encoder-decoder for both reconstruction and segmentation. The encoder architecture consists of $10$ convolutional layers, and the decoder contains $10$ layers of convolutions with up-sampling and $4$ additional convolutional layers. The network has been trained in an end-to-end fashion using a combined loss function including a reconstruction loss, a dice loss, and binary cross entropy loss.

Different from most of the above literature, based on pre-trained or existing backbone models, in this work we have optimized the architecture of two different networks for CT classification: (i)~a generative model based on Conditional VAE~\citep{Sohnetal2015},~\citep{Ivanovetal2019} used for learning {\em disentangled representation}, and (ii)~a downstream CNN network for supervised classification. Both models take as input a dual-channel information, i.e., the masked lung images and their corresponding lesion regions segmentation masks, obtained from a  multi-class lesion segmentation model~\citep{Tilborghsetal2020}, as described in Section~\ref{sec:Diagnosis}.

\section{COVID-19 CT data-set}
\label{sec:Data}

We collected 1419 non-contrast enhanced CT scans covering the entire lungs, comprising 795 COVID-19 positive cases and 624 COVID-19 negative cases. Data was retrospectively collected from $10$ institutions throughout Europe and Latin-America during the current pandemic. Positive cases were all confirmed with at least one positive RT-PCR test. Inclusion criteria for negative cases included an initial negative RT-PCR test result, and confirmation through repeated RT-PCR testing over time and/or radiological confirmation of the absence of lesions or the presence of an alternative pathology. All scans collected in this way concerned symptomatic patients.

Data from $9$ centers was pooled and complemented with scans from the open LIDC-IDRI~\citep{Armatoetal2011} data and scans of lung pathologies acquired prior to November $2019$, and used for training, validation and test sets. Data from one center was kept aside and used as an independent test-set. Table~\ref{tab:data} summarizes the collected data used for this study.

\begin{center}
	\begin{figure}[thp!]
		\centering
		\includegraphics[width=1\columnwidth, keepaspectratio]{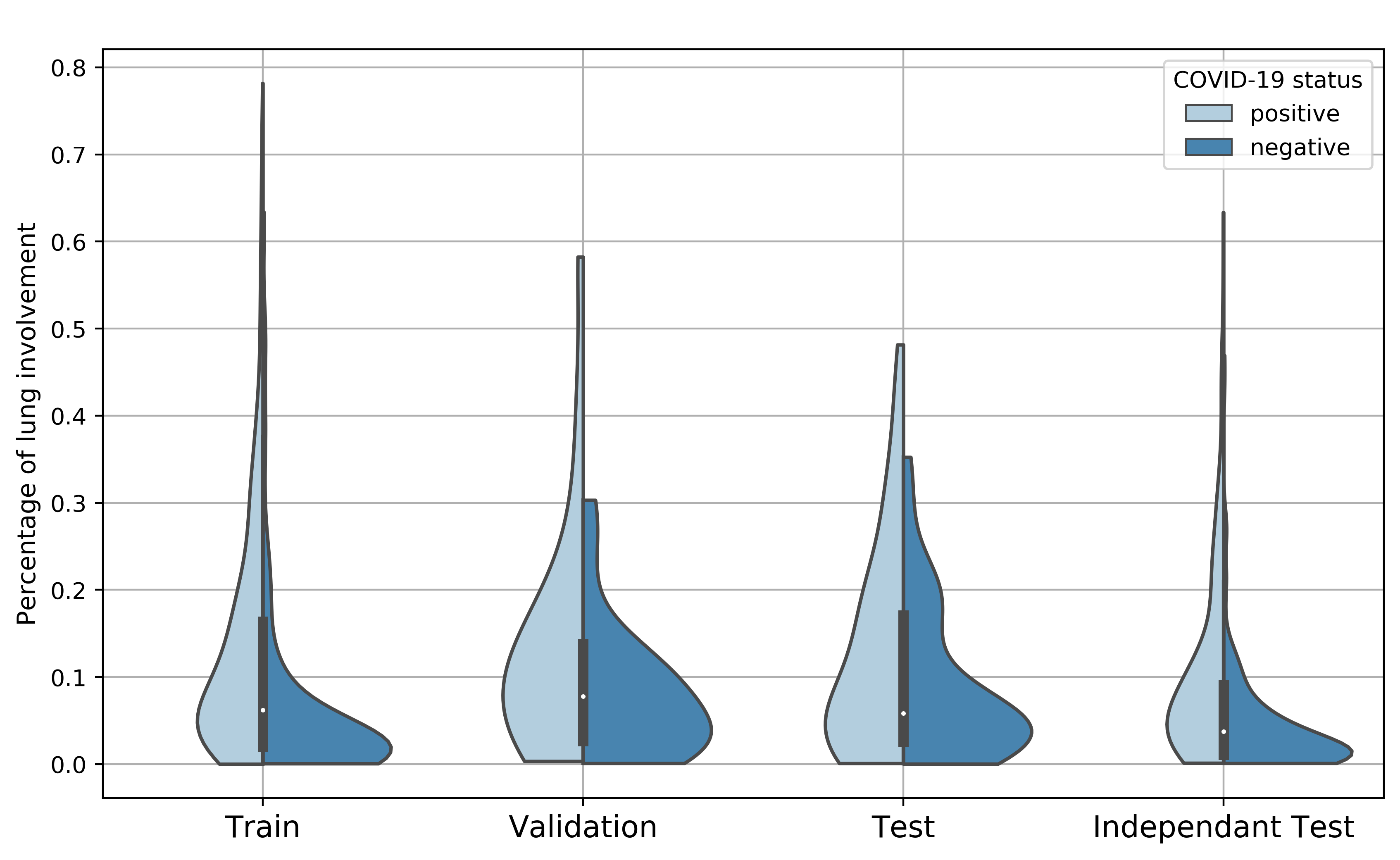}
		\caption{Percentage of lung involvement for the training, validation, test, independent-test sets per COVID-19 status.}
		\label{fig:lung_involvement_set_res}
	\end{figure}
\end{center}

\begin{center}
	\begin{figure}[thp!]
		\centering
		\includegraphics[width=1\columnwidth, keepaspectratio]{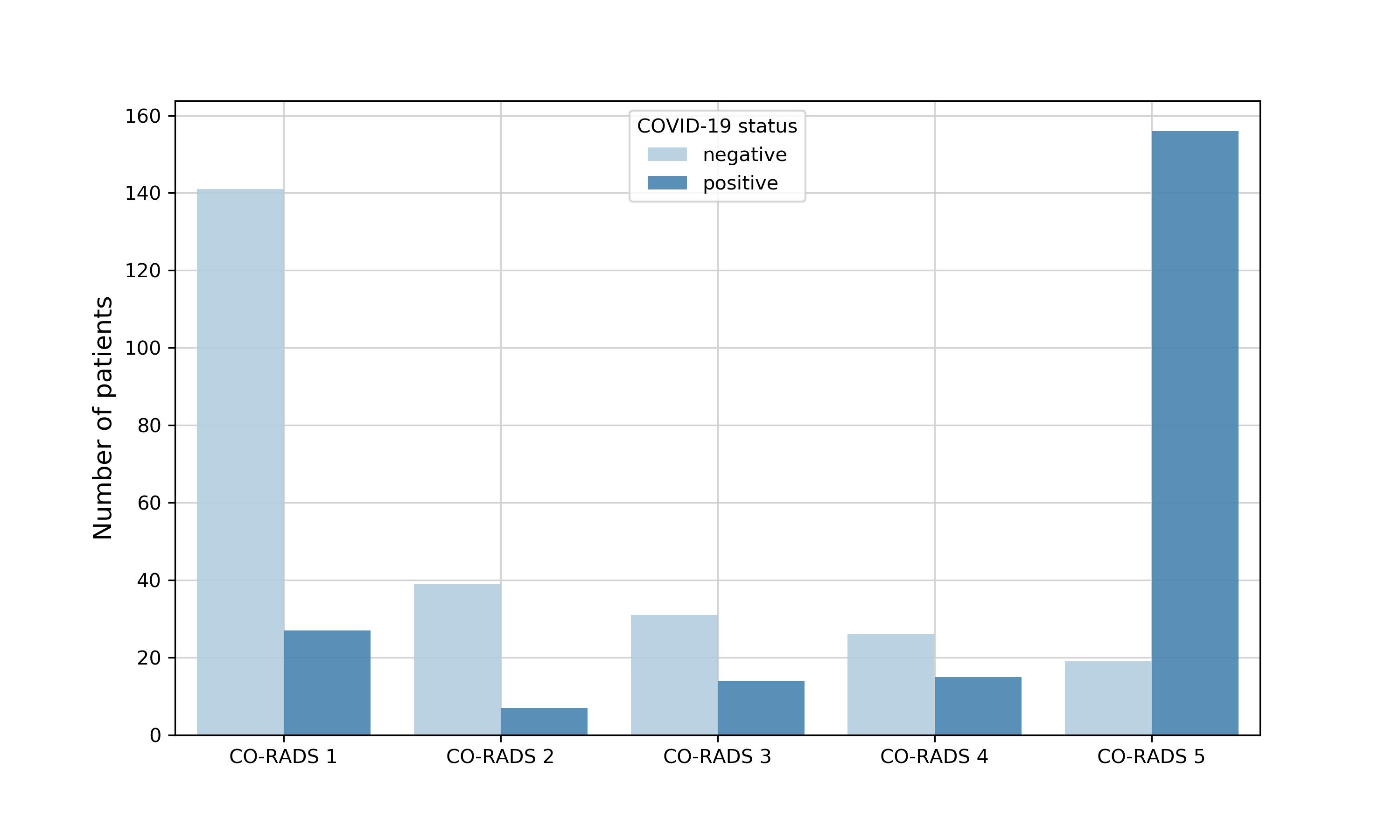}
		\caption{Number of patients per CO-RADS classification and COVID-19 status.}
		\label{fig:CORADS_occurencies}
	\end{figure}
\end{center}

The characteristics of the data was analysed in terms of percentage of lung involvement. To this end, all pathological tissue was delineated using the binary lesion segmentation method \textit{2DS} described in~\citep{Tilborghsetal2020}. This consists of a 2D U-Net~\citep{Ronnebergeretal2015} with $5$ levels and $16$ features at the first level. The model samples patches of $128 \times 128$ from the axial plane within the bounding box around the lungs. The percentage of lung involvement is then derived as the volume of the affected tissue over the total lung volume. The ratio of lung involvement is shown in Figure~\ref{fig:lung_involvement_set_res}. Additionally, the independent test-set was rated according to CO-RADS assessment scheme~\citep{Prokopetal2020} (Figure~\ref{fig:CORADS_occurencies}).

\begin{center}
\begin{table*}
\caption{Data description.}
\label{tab:data}
\centering
\begin{tabularx} {0.9\textwidth}{r r X X X r r}
\toprule													
	&	\# Scans	&	COVID-19 status	&	Source of data	&	Pathology	\\
\midrule													
Training 	&	150	&	Negative	&	LIDC-IDRI	&	NSCLC	\\
&	143	&	Negative	&	Pooled Centers	&	Various			\\
&	504	&	Positive	&	Pooled Centers	&	COVID-19		\\
\addlinespace													
Validation	&	0	&	Negative	&	LIDC-IDRI	&	NSCLC		\\
&	25	&	Negative	&	Pooled Centers	&	Various		\\
&	24	&	Positive	&	Pooled Centers	&	COVID-19		\\
\addlinespace													
Test	&	0	&	Negative	&	LIDC-ICRI	&	NSCLC	\\
&	50	&	Negative	&	Pooled Centers	&	Various	\\
&	48	&	Positive	&	Pooled Centers	&	COVID-19	\\
\addlinespace													
Independent	&	256	&	Negative	&	Independent Center	&	Various		\\
test	&	219	&	Positive	&	Independent Center	&	COVID-19	\\
\bottomrule																																						 
\end{tabularx}
\end{table*}
\end{center}

\section{CT-Based COVID-19 Diagnostic System}
\label{sec:Diagnosis}

\subsection{System Overview}

The pre-processing part of our workflow starts with lung segmentation, to outline lungs as region of interest for all following steps, for which the chest CT images intensities were clipped at $-1100$ HU and $100$ HU and rescaled to the $[0, 1]$ range. Automatic lung segmentation is performed using a 3D neural network based on the DeepMedic architecture~\citep{Kamnitsasetal2016} available in the consortium~\citep{Tilborghsetal2020}.

The second step of the processing pipeline consists of a multi-class lesion segmentation approach, from which we extract the lesion segmentation masks. We use a pre-trained multi-class lesion segmentation approach~\citep{Tilborghsetal2020} to identify the main patterns suggestive for COVID-19, namely ground glass opacities (GGO), consolidation (CO) and crazy paving pattern (CPP). In brief, the segmentation network is composed of the following elements: a 3D U-Net~\citep{Cicceketal2016} with $4$ levels, $32$ features at the lowest level, leaky rectified linear unit (ReLU), instance normalization~\citep{Ulyanovetal2016}, Max Pooling, and linear up-sampling. The generalized Wasserstein Dice loss~\citep{Fidonetal2017} is employed to deal with hierarchical ground truth labels. In this case, the label \textit{combined pattern}, was used for any combination of $2$ or more of the above-mentioned patterns (GGO, CO, CPP) for regions where reproducible separation of the individual patterns was not feasible during manual annotation. We refer the reader to~\citep{Tilborghsetal2020} for a detailed description of the preprocessing, training, and analysis of comparative performance with respect to other segmentation approaches in the field.

Both the masked lung images and their corresponding lesion segmentation masks are input to the following steps of the diagnostic system, i.e., representation learning module, CT classification module, and explainability module, for which the chest CT images intensities were cropped at $[-1000, 400]$ HU in order to preserve brighter structures within the lungs, and re-sampled with variable in-plane resolution and $3$~mm slice increment.

\subsection{Classification Model}
\label{sec:classification_model}

Image classification is the task of assigning a class label from a fixed set of categories to a given input image, for which deep learning models have shown great promise. Deep learning methods are representation-learning methods with multiple levels of representation layers. In the context of CNN, the feature representation learning task consists of optimizing the parameters of the CNNs, given the training data, to discover the underlying properties~\citep{Donahueetal2014, Sharifetal2014}, but this representation is heavily dependent on the data-set used for training.

In the case of COVID-19, considering the diversity of CT findings associated with COVID-19, a positive patient can show different types of lesions and levels of severity of infection as the illness evolves~\citep{Chungetal2020}. As we wish to learn a representation that is faithful to the underlying data structure, we followed the approach of {\em disentangled representation}, suggested in the seminal work of Bengio {\em et al.}~\citep{Bengioetal2013}, where different high-level generative factors are independently encoded, thus capturing the multiple explanatory factors and sharing factors across tasks priors. In our case, a high-level representation of COVID-19 could include bilateral and peripheral ground-glass and consolidative pulmonary opacities, sometimes with a rounded morphology and peripheral lung distribution~\citep{Bernheimetal2020}. Learning such disentangled factors of variation should allow the model to generalize more easily to similar, but different (from different institutions/cases), data.

In the literature, three main paradigms address disentangled learning~\citep{Higginsetal2018},~\citep{Ruizetal2019}: unsupervised, supervised, and weakly-supervised learning. Unsupervised models~\citep{Ovenekeetal2016} are trained without specific information about the generative factors of interest. They can identify simple explanatory components, however, do not allow latent variables to model specific high-level factors. Supervised models are learned by using a training set where the factors of interest are explicitly labeled. Following this paradigm, we can find different models, such as the semi-supervised deep generative models~\citep{Kingmaetal2014}, and the conditional variational auto-encoders~\citep{Sohnetal2015}. Weakly-supervised approaches use implicit information about factors of variation provided during training, such as the reference-based variational autoencoders~\citep{Ruizetal2019}, able to impose high-level semantics into the latent variables by exploiting the weak supervision provided by the reference set.

In the past years, variational autoencoder (VAE)~\citep{KingmaWelling2014}, coupling deep learning with variational inference, have emerged as a powerful latent variable model able to learn abstract data representations. VAE includes an encoder that transforms a given input into a typically lower-dimensional representation, and a decoder that reconstructs the input based on the latent representation. VAE learns an efficient feature representation that can be used to improve the performance of any standard supervised learning algorithm.

VAE learns a distribution over objects $p(x)$ and allows sampling from this distribution. In our case, we are interested in learning a conditional distribution $p(x|y)$, where $x$ is a CT chest image, and $y$ could be the characteristics describing the affected COVID-19 lung (lesion's shape and appearance, class membership, etc…). Such CVAE~\citep{Sohnetal2015},~\citep{Ivanovetal2019} are popular approaches. In this work we use a generative model based on CVAE, relying on two inputs: the CT images and the class labels, and propose a representation learning algorithm that incorporates side information in the form of an additional label image (representing the segmented lung lesions) at training time to produce a more informed representation learning. Once we learn the CVAE encoder that maps images into a disentangled latent space, the weights of the encoder are used to initialize a downstream CNN network for supervised classification. Feature aggregation layers are added on top of the encoder base, and the resulting network is trained, in conjunction with a focal loss function, to classify CT chest image. With the explainable classification results, our system largely simplifies and accelerates the diagnosis process for manual reading.

\subsubsection{Conditional Variational Autoencoder}
\label{sec:conditional_variational_autoencoder}

\begin{center}
	\begin{figure*}[t!]
		\centering
		\includegraphics[width=0.9\textwidth, keepaspectratio]{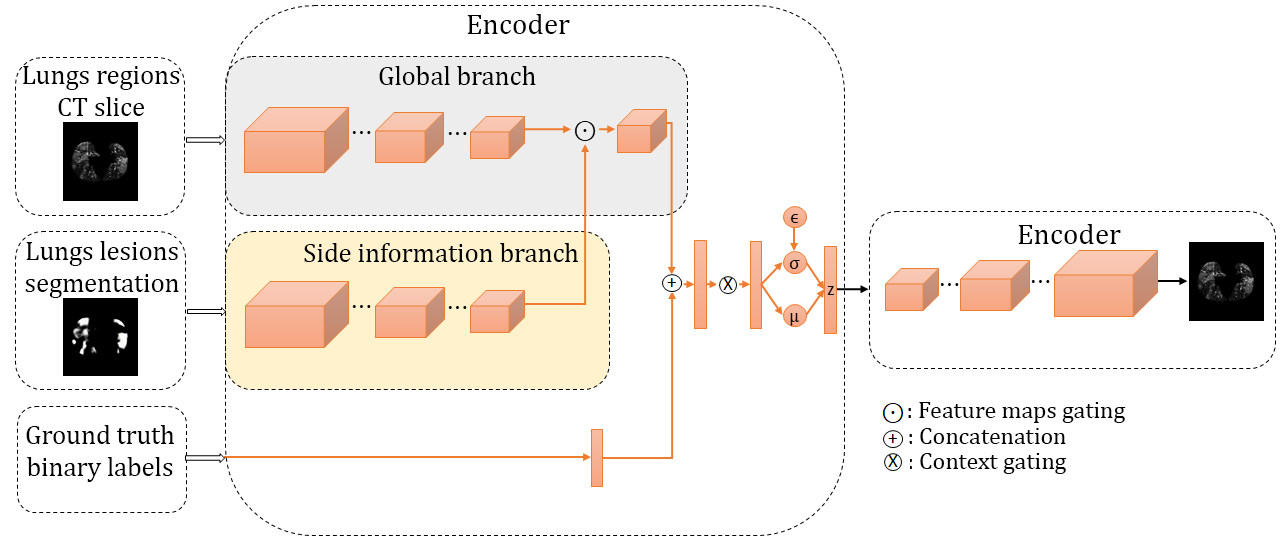}
		\caption{CVAE model details.  $\boldsymbol{\mu}, \boldsymbol{\sigma}$ are the Gaussian parameters and  $\boldsymbol{\epsilon}$ is auxiliary Gaussian randomness}
		\label{fig:CVAE}
	\end{figure*}
\end{center}

To outline the generative approach of the CVAE in our work, let's define $x$ as a training data instance corresponding to a certain patient and $y$ the binary ground truth labels (COVID-19 positive or COVID-19 negative) for such a patient. The objective of CVAE is to approximate the conditional variable distribution $p(\hat{x}|x,y)$, $x$ and $y$ being the primary input to the CVAE and $\hat{x}$ the output. Then, the stochastic latent variable $z$ is drawn
from the prior distribution $p(z|x,y)$ and the data instance $\hat{x}$ is generated from the generative distribution $p(\hat{x}|z)$. As such, the generative distribution is conditioned on the data instance and the binary label corresponding to the patient. CVAE models the distributions using Neural Networks to learn their parameters, formally defined as follows:

\begin{equation}
\begin{gathered}
p_{\theta} (z|x,y) = \varphi_{prior}(x,y;\theta)\\
q_{\phi} (z|x,y) = \varphi_{cond}(x,y;\phi)\\
p_{\psi} (\hat{x}|z) = \varphi_{gen}(z;\psi)\\
\end{gathered}
\end{equation}

\noindent where $p_{\theta} (z|x,y)$, $q_{\phi}(z|x,y)$ and $p_{\psi} (\hat{x}|z)$ are the prior, conditional and generative distributions respectively. The conditional $q_{\phi}(z|x,y)$ is an approximation of the posterior. We use Gaussian distributions, $\mathcal{N}(\boldsymbol{\mu}; diag(\boldsymbol{\sigma})^2)$, parameterized with means, $\boldsymbol{\mu}{_{pri}}$, $\boldsymbol{\mu}{_{cond}}$, and $\boldsymbol{\mu}{_{gen}}$, as well as standard deviations, $\boldsymbol{\sigma}{_{pri}}$, $\boldsymbol{\sigma}{_{cond}}$ and  $\boldsymbol{\sigma}{_{gen}}$, which are the outputs of the neural networks $\varphi_{prior}(x,y;\theta)$, $\varphi_{cond}(x,y;\phi)$ and $\varphi_{gen}(z;\psi)$ with learning parameters $\theta$, $\phi$ and $\psi$ respectively.

A general overview of the proposed CVAE-based disentangled representation learning is depicted in Figure~\ref{fig:CVAE}. It involves one encoder with two branches, the latent variables and the decoder. The encoder consists of a global branch along with side information branch, each one consists in a CNN. Furthermore, the CNN in the side information branch is symmetric to the CNN of the global branch, namely it has an identical set of filters. The difference between them is that each CNN accepts distinct input. The input to the global branch are the CT images within the CT chest volume corresponding to a certain patient, and the input to the side information branch is the lesion segmentation obtained from such CT images, that is, both inputs are paired. In this manner, our CVAE learns the generative distribution conditioned also on relevant lesions related to COVID-19.

The architecture of the global and side information CNN branches incorporates one convolutional layer of $16$ $5 \times 5$ filters with a ReLU layer as activation function followed by a max pooling (MP) layer. Subsequently, a second convolutional layer of $32$ $3 \times 3$ filters is employed, with leaky ReLU layer as activation function followed by MP. Afterwards, a third convolutional layer of $64$ $3 \times 3$ filters, also with leaky ReLU layer as activation function. The two CNN branches in the encoder operate simultaneously. Consequently, after the third convolutional layer we obtain two feature maps of the same dimensions, that is, the feature map from the global CNN branch and a ``relevance'' map from the side information CNN branch.

The input to the side information CNN is the (binary) lesion segmentation mask. We wish to enforce high activation in these regions of interest for diagnosis. We therefore apply a gating mechanism in which the resulting feature maps from the global and side information CNN branches are multiplied in an element-wise manner to enforce the attention to such a regions. Subsequently, a ReLU is applied to the attained feature maps to enable information flow while preventing the gradients to vanish. The gating mechanism is formally defined as follows:
\begin{equation}
\label{equGating_1}
\begin{gathered}
F =g(F_{g_{cnn}}\odot F_{s_{cnn}})
\end{gathered}
\end{equation}
\noindent where $F_{g_{cnn}}$ and $F_{s_{cnn}}$ are the feature maps resulting from the global and side information branches respectively, $\odot$ is the Hadamard product and $g$ the ReLU. The proposed gating approach acts as an inhibitor of the feature maps from the global branch activated on those regions where the salient information is not related to COVID-19 findings. In other words, the gating mechanism allows selecting relevant visual information for COVID-19 diagnosis. Hence, it empowers the model to focus attention on the most meaningful visual cues for such a task, which leads to purposeful feature representations.

To condition the VAE to the input ground truth labels, we feed the ground truth labels into an embedding neural network to attain fixed-length low dimensional representation:

\begin{equation}
\begin{gathered}
\label{equInputEmb}
e = \varphi_{embedding}(y;\theta)
\end{gathered}
\end{equation}

\noindent where $y$ is the input binary ground truth label and $\varphi_{embedding}$ is a Neural Network  with learning parameters $\theta$.

Once the low-dimensional representation of the binary ground truth labels is computed, we reshape the feature maps obtained in Equation~\ref{equGating_1} and pass the reshaped vector into a neural network. Afterwards, the output of the neural network is concatenated with the representation of the binary ground truth to merge thereafter using a neural network. As a result, we expect attaining a representation that synthesizes and carries purposeful visual information from the input data instance, due to the gating with the corresponding lesion segmentation, whereas enforcing the prior to consider the binary labels:

\begin{equation}
\begin{gathered}
\label{equPriorCond}
f = \varphi_{merge}(\varphi([F,e]);\theta)
\end{gathered}
\end{equation}

\noindent where $\varphi$ and $\varphi_{merge}$ are neural networks with non-linear activation and learning parameters $\theta$. The operator $[ \cdot ]$ represents a concatenation between the reshaped feature maps and the ground truth label embedding.

The components of the representation $f$ contain salient information from the lungs and lesion input $x$ and the binary label $y$. Therefore, we propose to model non-linear inter-dependencies between these feature representations using a context gating (CG) mechanism. Indeed, the role of the CG mechanism is to re-calibrate the significance of each dimension of the representation $f$. For the CG mechanism, we use a neural network followed by a non-linear activation and the output of the neural network is multiplied element-wise by the input representation, formally~\citep{Kmiecetal2017}:

\begin{equation}
\begin{gathered}
\label{equContextGating_2}
\hat{f}=\varphi_{cg}(f;\theta) \odot f
\end{gathered}
\end{equation}

\noindent where $f$ is the input vector representation, $\varphi_{cg}$ is a neural network with a non-linear activation function and learning parameters $\theta$, $\hat{f}$ is the transformed output vector.


The architecture of the decoder in the CVAE consists of one transposed convolutional layer of $32$ $3 \times 3$ filters with Leaky ReLU layer as activation function followed by bilinear interpolation for up-sampling. Subsequently,  a second transposed convolutional layer is employed of $16$ $3 \times 3$ filters, with Leaky ReLU layer as activation function followed by bilinear interpolation for up-sampling. Finally, a third transposed convolutional layer of one $2 \times 2$ filters, also with Leaky ReLU layer as activation function and followed by bilinear interpolation for up-sampling.


To learn the parameters of the CVAE we optimize the variational lower bound, that is an appropriate choice for the problem addressed~\citep{Sohnetal2015}. Therefore, we introduce the following objective function:

\begin{equation}
\label{equLossFunction}
\begin{gathered}
\resizebox{0.9\hsize}{!}{$L_{CVAE} (x,y,\theta,\phi,\psi)=\mathbb{E}_{q_{\phi}(z|x,y)} \left[\log p_{\psi} (\hat{x}|z) \right]- D_{KL} (q_{\phi} (z|x,y)||p_{\theta} (z|x,y))$}
\end{gathered}
\end{equation}

\noindent where $D_{KL} (q_{\phi} (z|x,y)||p_{\theta} (z|x,y))$ is the Kullback–Leibler divergence, $\theta$, $\phi$, $\psi$ are the model trainable parameters. We use Gaussian distributions to model the latent variables and apply the parametrization approach of Kingma and Welling~\citep{KingmaWelling2014}. The basic idea is to approximate the sampling with auxiliary Gaussian randomness $\boldsymbol{\epsilon}$. Then, for each Gaussian distribution $\boldsymbol{\epsilon}$ is element-wise multiplied by the standard deviation, and the resulting value is added to the mean as follows: $\boldsymbol{\mu}+\boldsymbol{\sigma}\odot \boldsymbol{\epsilon}$. In such a way, we do not sample directly from the Gaussian distribution and we can optimize the parameters of the CVAE using Stochastic Gradient Descent.

\subsubsection{Pairing the Encoder for Classification}
\label{sec:Pairing_CNN}

\begin{center}
	\begin{figure*}[t!]
		\centering
		\includegraphics[width=0.9\textwidth, keepaspectratio]{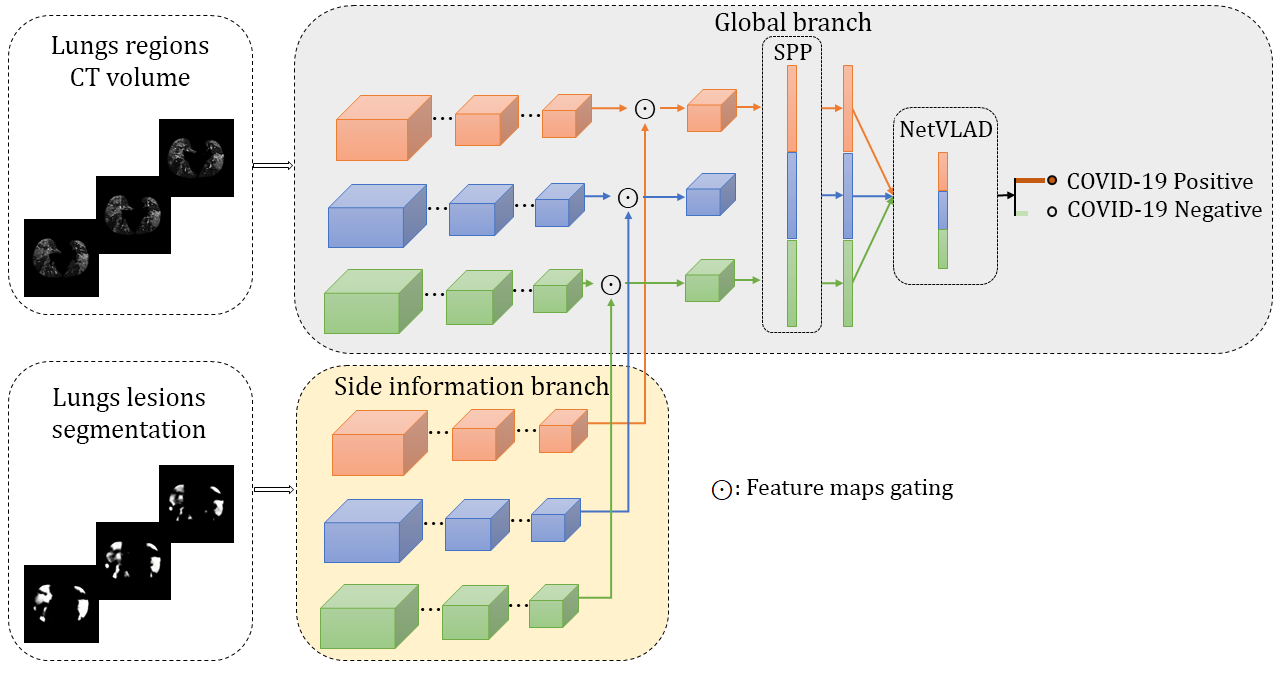}
		\caption{Classification framework. It takes as input a series of CT slices and generates a CT volume classification prediction. The SPP features maps of CT slices are fed to a NN followed by the NetVLAD layer and the resulting feature map is fed to a fully connected layer to generate a probability score for each class.}
		\label{fig:Classification}
	\end{figure*}
\end{center}

For the final COVID-19 diagnosis task, we build upon the parameters of the encoder optimized on the CVAE as illustrated in Figure~\ref{fig:Classification}. We therefore transfer the learned parameters to a classification model for the final COVID-19 diagnosis with the objective of taking advantage of the parameters already trained using the same global and side information branches. Hence, our COVID-19 classification model has exactly the same structure as the encoder layers used in the CVAE.

As can be observed from Figure~\ref{fig:Classification}, both CNN branches as well as the gating mechanism between them are identical to the CVAE encoder, to explicitly force the attention of the model on the most meaningful visual information for COVID-19 diagnosis.  However, in contrast to the CVAE we do not use labels as input. Furthermore, we incorporate a spatial pyramid pooling (SPP)~\citep{HeKetal2015} layer, replacing the third max pooling (MP) layer of the encoder of the CVAE. The SPP layer is followed by a neural network. Subsequently, we adopt the NetVLAD~\citep{Arandjelovicetal2016} approach to aggregate CT slice-level features over the whole CT volume. Finally we employ a linear layer with $2$ neurons for the diagnosis score.

The SPP layer is suitable for COVID-19 diagnosis because it can capture features at multi-level spatial bins and at different scales, while the features are computed over the whole slice at once. Additionally, the SPP layer increases the capacity of our model to deal with variable size inputs as it outputs fixed-length representations regardless of input size, enabling further benefits, such as the possibility of training with variable size images to enhances the scale-invariant properties.

Different from most previous work (see Section~\ref{sec:Related Work}) using $3D$-CNNs or $2D$-CNNs in conjunction with pooling mechanisms to classify a CT volume, in our model we adopt the NetVLAD layer~\citep{Arandjelovicetal2016}, with $M$ clusters, to aggregate the slice-level descriptors, provided by the SPP layer, for classification. NetVLAD offers a powerful aggregation mechanism with learnable parameters that can be easily plugged into any CNN architecture.

The focal loss~\citep{Linetal2017b} ($FL_{CM}$) was chosen over the widely used binary cross entropy loss as the latter has a tendency to be influenced heavily by trivial examples in case of class imbalance. Moreover focal loss increases the focus on hard misclassified training data instances:
%

\begin{equation}
FL_{CM}=-\lambda_t (1-\hat y_t)^\gamma \log(\hat y_t)
\end{equation}
\noindent where $\gamma$ and $\lambda_t$ are adjustable parameters to deal with hard misclassified samples and class imbalance respectively, and
\begin{equation*}
\begin{gathered}
\hat y_t = \left\{\begin{array}{l}	\hat{y} \begin{array}{c} \quad \quad \quad if \quad y=1,	\end{array}\\
									1 -\hat{y} \quad \begin{array}{c} otherwise. \end{array} \end{array} \right.\\
\end{gathered}
\end{equation*}

\subsection{Explainable Model}

We provide visual explanations for the network decision in the form of heatmaps that highlight the excitatory features captured by the model during the forward prediction pass. We apply the layer-wise relevance propagation (LRP) framework proposed by Bach {\em et al.}~\citep{Bachetal2015}, where individual pixel contributions or \textit{relevance} scores are computed by back-propagating the activation of the class of interest to the input.  Our approach is based on the composite strategy $LRP_{CMP}$~\citep{Montavonetal2019}, which consists of applying different LRP propagation rules at different stages of the network. We employ the $LRP_0$ rule at the classifier stage. For a linear layer, $LRP_0$ redistributes relevance scores $R_j^{(l+1)}$ at layer $l+1$ to the preceding layer $l$ as:
\begin{equation}
R_i^{(l)} = \sum_j\frac{r_{ij}}{r_j}R_j^{(l+1)},
\end{equation}
where $r_{ij} \equiv x_i w_{ij}$, $r_j \equiv \sum_i r_{ij}$, with $x_i$ corresponding to the forward pass activations of layer $l$ and $w_{ij}$ being the layer weights. This simple rule is similar to the \textit{Gradient~$\times$~Input} method~\citep{Shrikumaretal2016}. For the feature extraction stage (i.e., convolutional layers), we employ the $LRP_{\alpha \beta}$ that treats positive and negative contributions asymmetrically:
\begin{equation}
R_i^{(l)} = \sum_j\left(\alpha\frac{r_{ij}^+}{r_j^+} + \beta\frac{r_{ij}^-}{r_j^-}\right)R_j^{(l+1)},
\end{equation}
with $\alpha + \beta = 1$, $r_{ij}^+$ and $r_{ij}^-$ corresponding to the positive and negative preactivations respectively. This rule leads to heatmaps that are less noisy, as it is more robust to gradient shattering. In our case, we use $LRP_{\alpha \beta}$ with $\alpha=2$ and $\beta=-1$. Extending these rules to convolutional layers is straightforward. For the max pooling layers, including the spatial pyramidal pooling stage, we only propagate relevance to the maximum activation path. For the image-level layer, we apply the $LRP_{z^B}$ rule~\citep{Montavonetal2019}.

Precautions must be taken for the attention (gating) and the aggregation stages (NetVLAD), for which there exists no well-defined LRP rules. In these cases, we use the \textit{Gradient~$\times$~Input} model~\citep{Shrikumaretal2016} (similar to $LRP_0$ in essence) to back-propagate the incoming relevance scores. Moreover, during the backward pass, we approximate the feature aggregation step of NetVLAD by a linear matrix multiplication by freezing the clustering matrix, so that explanations are limited to features important to the last classification stage.

\section{Experimental results}
\label{sec:Experiments}

\subsection{Implementation and training details
\label{sec:implement_and_data-sets}}
We have implemented the proposed CVAE and classification model on top of Pytorch\footnote{\url{https://pytorch.org/}}. The neural networks used in the CVAE for embedding the binary ground truth labels and merging the embedding with the gated feature maps have $64$ neurons. The number of neurons of the neural network used for the context gating mechanism is set to $64$. Furthermore, the dimension of the latent variables was set to $16$. That is to say, the neural networks employed to parameterize the Gaussian distributions have $16$ neurons. We trained the architecture illustrated in Figure~\ref{fig:CVAE} using Adam for $200$ epochs with an initial learning rate of $5e-3$.
For the classifier (Figure~\ref{fig:Classification}), the SPP layer has three MP levels whose kernels' sizes are $5 \times 5$, $3 \times 3$, and $2 \times 2$ respectively. The neural network following the SPP has $512$ neurons. NetVLAD is employed with $M=64$ clusters. The classification model was also trained using Adam for $200$ epochs and initial learning rate of $1e-5$. Additionally, we regularized the model with a weight decay of $1e-5$ and early stopping. The focal loss parameter $\lambda$ was set to $0.25$ and $0.35$ for the classes COVID-19 Negative and COVID-19 Positive respectively, and the parameter $\gamma$  was set to $5$. We built upon the validation set to empirically tune all the adjustable parameters and we did not employ any data augmentation technique. For the lung and lesion segmentation, we used the pre-trained models~\citep{Tilborghsetal2020}.

\subsection{Ablation Study}

We progressively designed the proposed model, by thoughtfully incorporating the different components, following the data provision from our partner hospitals. To evaluate the benefit of the proposed approach, we performed an ablation study on a small data-set (Table~\ref{tab:data_small}), consisting of $122$ COVID-19 negatives and $74$ COVID-19-positives, provided late March 2020 by two hospitals. Table~\ref{table:AblationAll} summarizes the ablation study results. As can be seen, the classification model without CVAE and side information provided a poor classification performance. This confirmed the poor feature representation obtained by a simple CNN. Hence, aiming at learning a disentangled representation, we incorporated the CVAE with the global branch. One can notice that once the CVAE was integrated and paired with the classification model the performance increased significantly; reaching higher scores over all the reported metrics. This confirms that coupling the variational inference to learn faithful abstract representations is an appropriate approach to improve the performance of supervised learning algorithms. The performance is increased further when the side branch information is incorporated. We hypothesise that the feature maps gating contributed to enhance the attention of the model on patterns suggestive of COVID-19, which improved the diagnosis.

\label{sec:ablationstudy}
\begin{table}[!h]
		\caption{Data employed for the ablation study.}
		\label{tab:data_small}
		\centering
		\resizebox{\columnwidth}{!}{\begin{tabular}{c c c c c}
			\toprule													
			&	\# Scans	&	COVID-19 status	&	Source of data	&	Pathology\\
			\midrule													
			Training &	75	&	Negative	&	Two Centers	&	Various\\
			&	48	&	Positive	&	Two Centers	&	COVID-19\\
			\addlinespace													
			Validation &	4	&	Negative	&	Two Centers	&	Various\\
			&	3	&	Positive	&	Two Centers	&	COVID-19\\
			\addlinespace													
			Test	 &	43	&	Negative	&	Two Centers	&	Various\\
			&	23	&	Positive	&	Two Centers	&	COVID-19\\
			\bottomrule													
		\end{tabular}}
\end{table}

\begin{center}
\begin{table}[htb]
\caption{Ablation study results. The ablation study was performed on the reduced data set described in Table~\ref{tab:data_small}.}
\label{table:AblationAll}
\centering
	\resizebox{\columnwidth}{!}{\begin{tabular}{rccc}
		\toprule
Model & Sensitivity & Specificity &$F_1$-score \\
\midrule
Classification without CVAE & 0.57 & 0.59   & 0.48  \\
CVAE without side information & 0.74 & 0.68 & 0.63    \\
Proposed model & 0.83 & 0.72   & 0.70  \\
		\bottomrule	
\end{tabular}}
\end{table}
\end{center}
\subsection{COVID-19 Diagnosis}

\begin{table}[h!]
	\caption{Final classification performance on the test and independent test set given in Table~\ref{tab:data}.}
	\label{table:exp2}
	\centering
	\resizebox{\columnwidth}{!}{\begin{tabular}{r c c c}
			\toprule
		&	\multicolumn{3}{c}{Evaluation metrics}\\
		&	Sensitivity & Specificity &$F_1$-score \\
			\midrule
	    Test-set (98 cases) &  0.94 & 0.82 & 0.88 \\
	    Independent test-set (475 cases) &  0.96 & 0.59 & 0.79 \\
			\bottomrule
	\end{tabular}}
\end{table}

\begin{center}
	\begin{figure}[thp!]
		\centering
		\includegraphics[width=1\columnwidth, keepaspectratio]{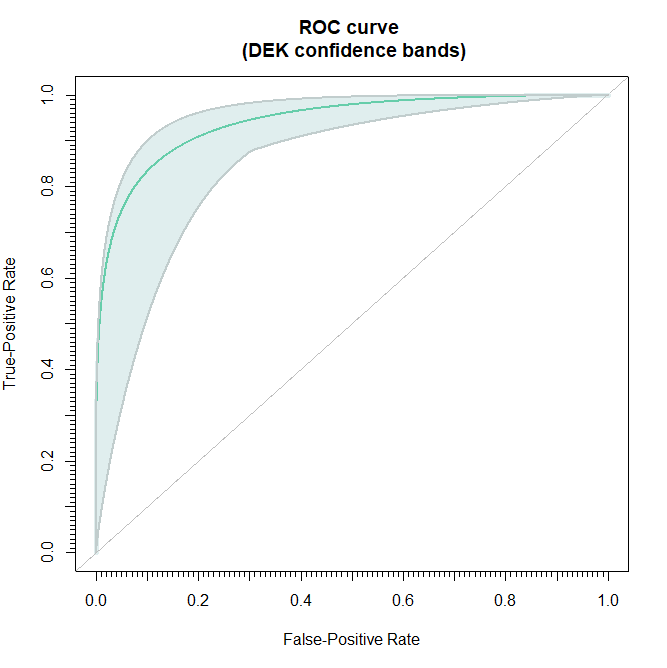}
		\caption{Receiver operating characteristic curve of the model, with $AUC=0.97 [0.93-0.99]$ for the test set of Table~\ref{tab:data}.}
		\label{fig:Exp3_ROC}
	\end{figure}
\end{center}

We conducted training, validation and testing on the data set described in Table~\ref{tab:data}. In Table~\ref{table:exp2} we report the outcomes of the proposed method. As can be seen our proposed COVID-19 diagnosis system  yielded satisfactory results. For the test-set, the proposed method diagnosed accurately $41$ COVID-19 Negative and $45$ COVID-19 Positive patients, respectively, while $12$ of the $98$ patients were misdiagnosed. The ROC curve is shown in Figure~\ref{fig:Exp3_ROC} and was generated by employing the bootstrapping method of Perez {\em et al.}~\citep{Perezetal2018} and the Demidenko confidence~\citep{Demidenko2012}, with $2000$ iterations and $95\%$ confidence interval. An AUC of 0.97 was obtained and the FPR for TPR=$95\%$ was 0.28[0.02-0.5].


\begin{center}
	\begin{figure}[thp!]
		\centering
		\includegraphics[width=1\columnwidth, keepaspectratio]{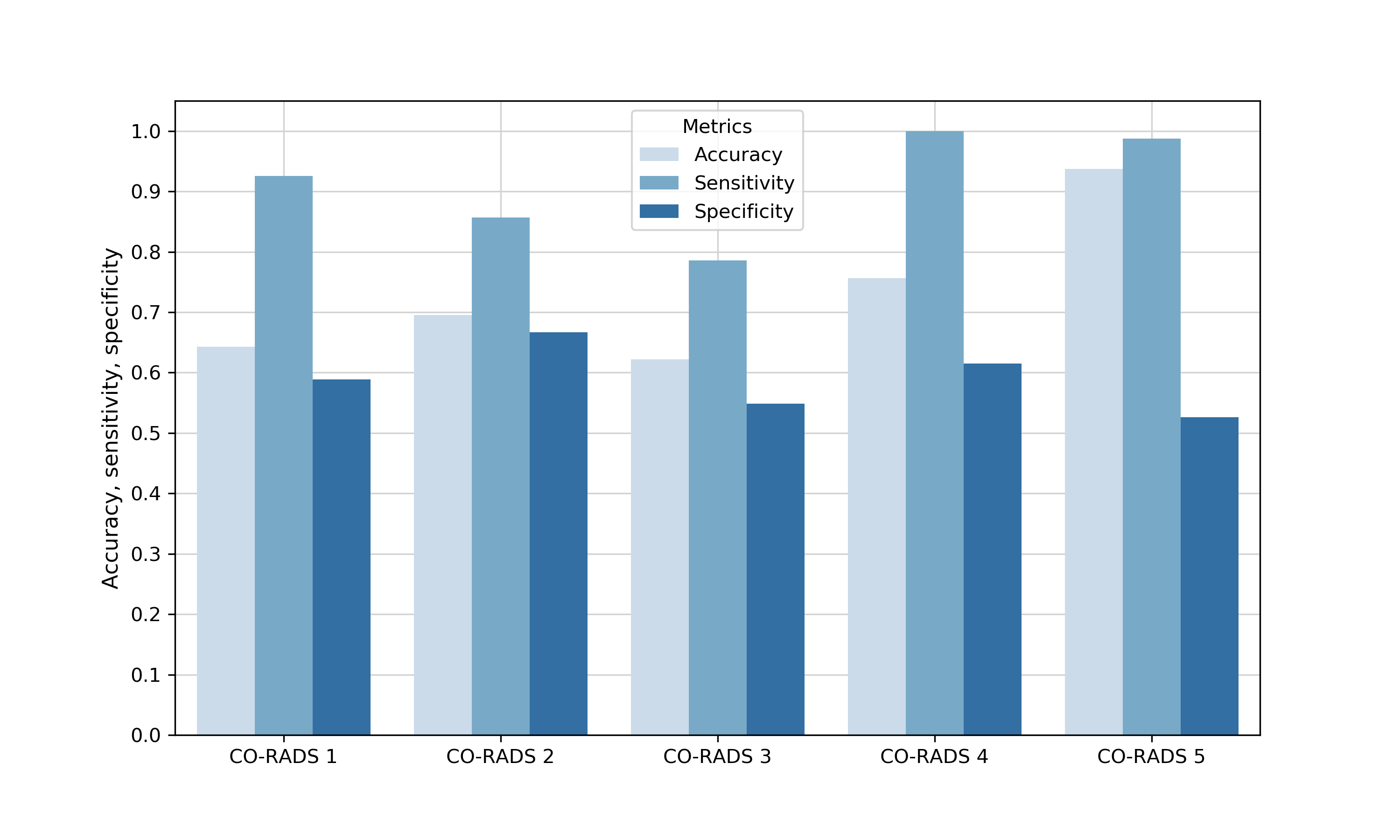}
		\caption{Accuracy, sensitivity and specificity per CO-RADS rating.}
		\label{fig:acc_sens_spc_per_CORADS}
	\end{figure}
\end{center}

Finally, aiming at validating the generalization of the proposed method, we conducted experiments on an independent test set of $475$ patients as listed in Table~\ref{tab:data}. We report the qualitative results on the independent test in Table~\ref{table:exp2}. On this challenging independent set, a lower specificity was obtained, while sensitivity was maintained. Accuracy, sensitivity and specificity were also computed per
CO-RADS rating (Figure~\ref{fig:acc_sens_spc_per_CORADS}). As expected, the achieved performance varied over the different CO-RADS classes. Surprisingly, the model was able to achieve high sensitivity for CO-RADS1 and CO-RADS2, accurately classifying 31 COVID-19 positive patients which were rated as very low and low level of suspicion for COVID-19 by radiologists, while missing only three.

\begin{center}
	\begin{figure}[thp!]
		\centering
		\includegraphics[width=1\columnwidth, keepaspectratio]{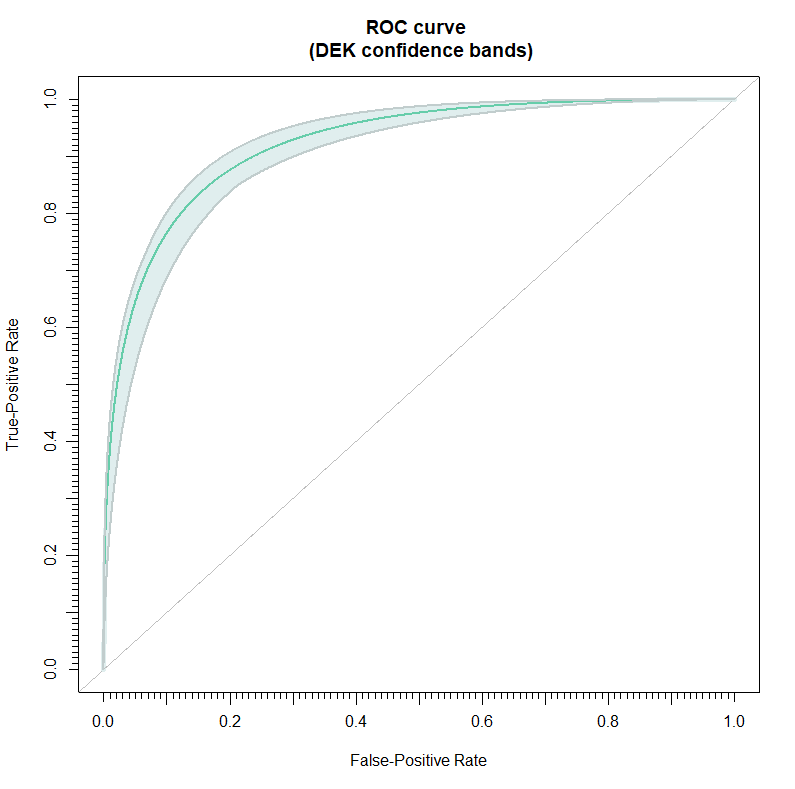}
		\caption{Receiver operating characteristic curve of the model on the independent data-set (475 patients), $AUC=0.93 [0.91-0.95]$.}
		\label{fig:Exp4_ROC}
	\end{figure}
\end{center}

The ROC curve is shown in Figure~\ref{fig:Exp4_ROC}, with an $AUC=0.93 [0.91-0.95]$ and the FPR for TPR=$95\%$ is $0.31[0.18-0.52]$.


%
%
%
%
%

\subsection{Visual explanations}

\begin{figure*}[h!] 
	\centering
	\includegraphics[width=0.42\linewidth]{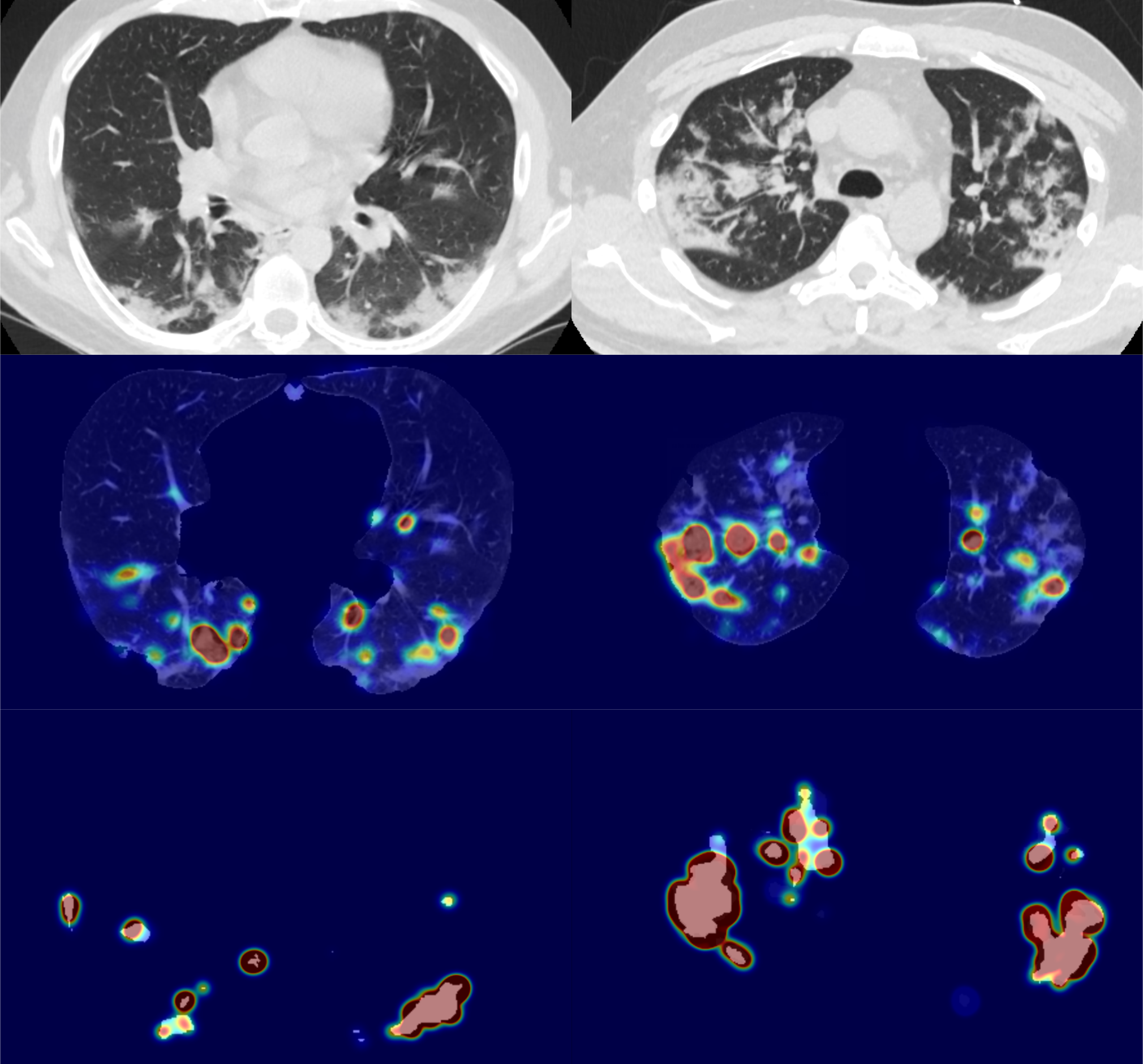}\hspace{-1mm}	 \includegraphics[width=0.42\linewidth]{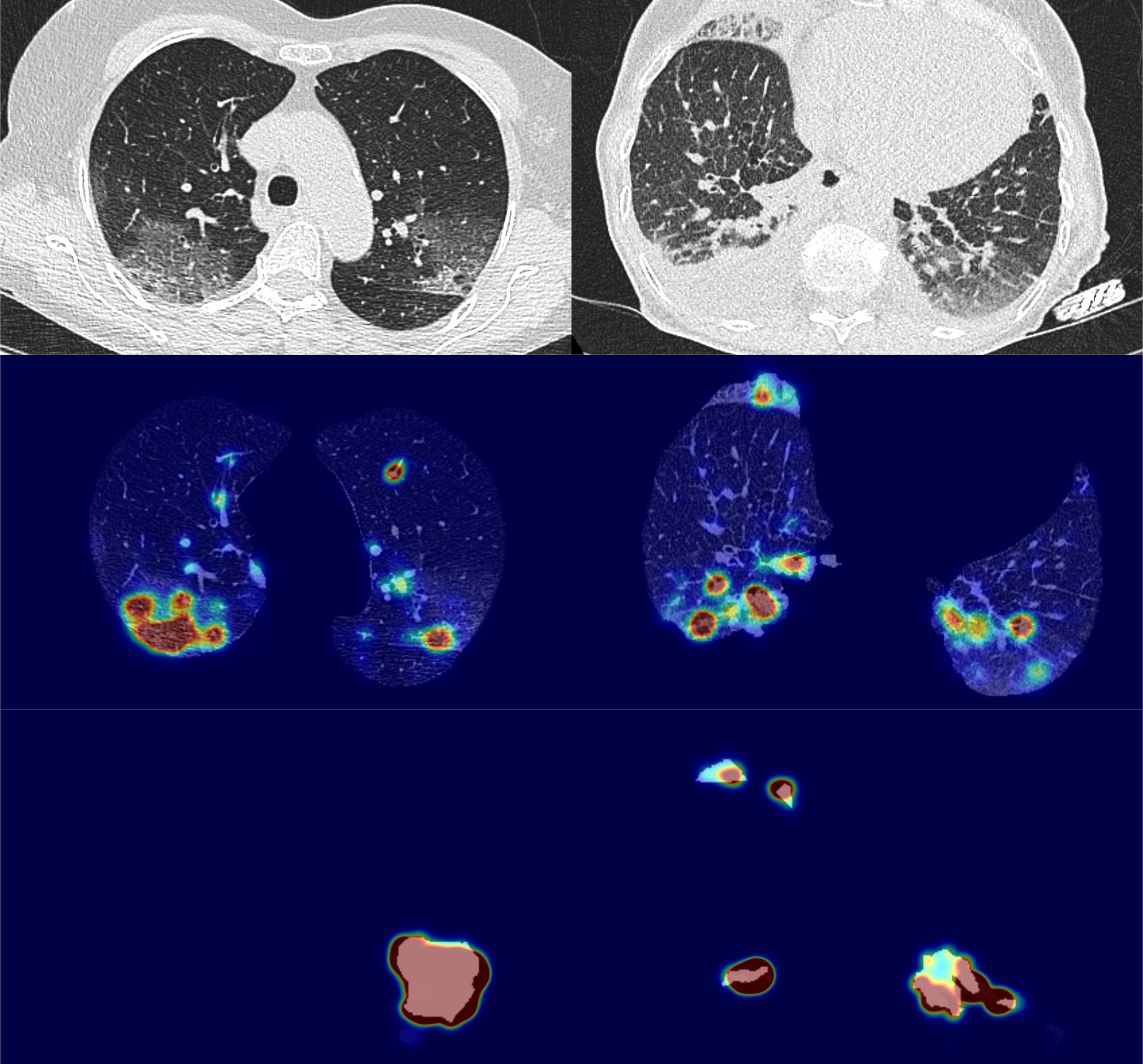}
	\caption{Examples of relevance heatmaps for correctly-classified COVID-19 positive patients (positive attributions only). From top to bottom: original CT-scan slice, segmented lungs input with relevance attributions, and lesions binary segmentation map with relevance attributions. Heatmaps are obtained using the $LRP_{CMP}$-based method and are post-processed using Gaussian filtering.}
	\label{fig:vis_explanations}
\end{figure*}

In Figure~\ref{fig:vis_explanations}, we provide 4 examples of relevance heatmaps for correctly-classified COVID-19 positive patients. Multi-focal consolidations and peripheral ground-glass opacities can be observed in the input images, and were mostly correctly segmented using the lesion segmentation process. We observe that features from both inputs are incorporated by the attention mechanism, with an important emphasis on the pre-identified lesions. The distribution of the heat on the segmented lungs either correlates with the segmented lesions, or appears to be imbalanced between the two input modalities. For instance, in the third example of Figure~\ref{fig:vis_explanations} where a ground glass region is left undetected by the lesions segmentation process. This demonstrates that the complementarity of both input features is correctly exploited. The lung heatmaps show that pulmonary abnormalities typical to COVID-19 (including peripheral consolidations) are brought out in combination to other normal structures; although the non-specific features are generally less highlighted. A careful inspection is still required to fully disentangle the different types of structures revealed by the explanation process.

\begin{center}
	\begin{figure}[h!] 
		\centering
		\includegraphics[width=1\columnwidth, keepaspectratio]{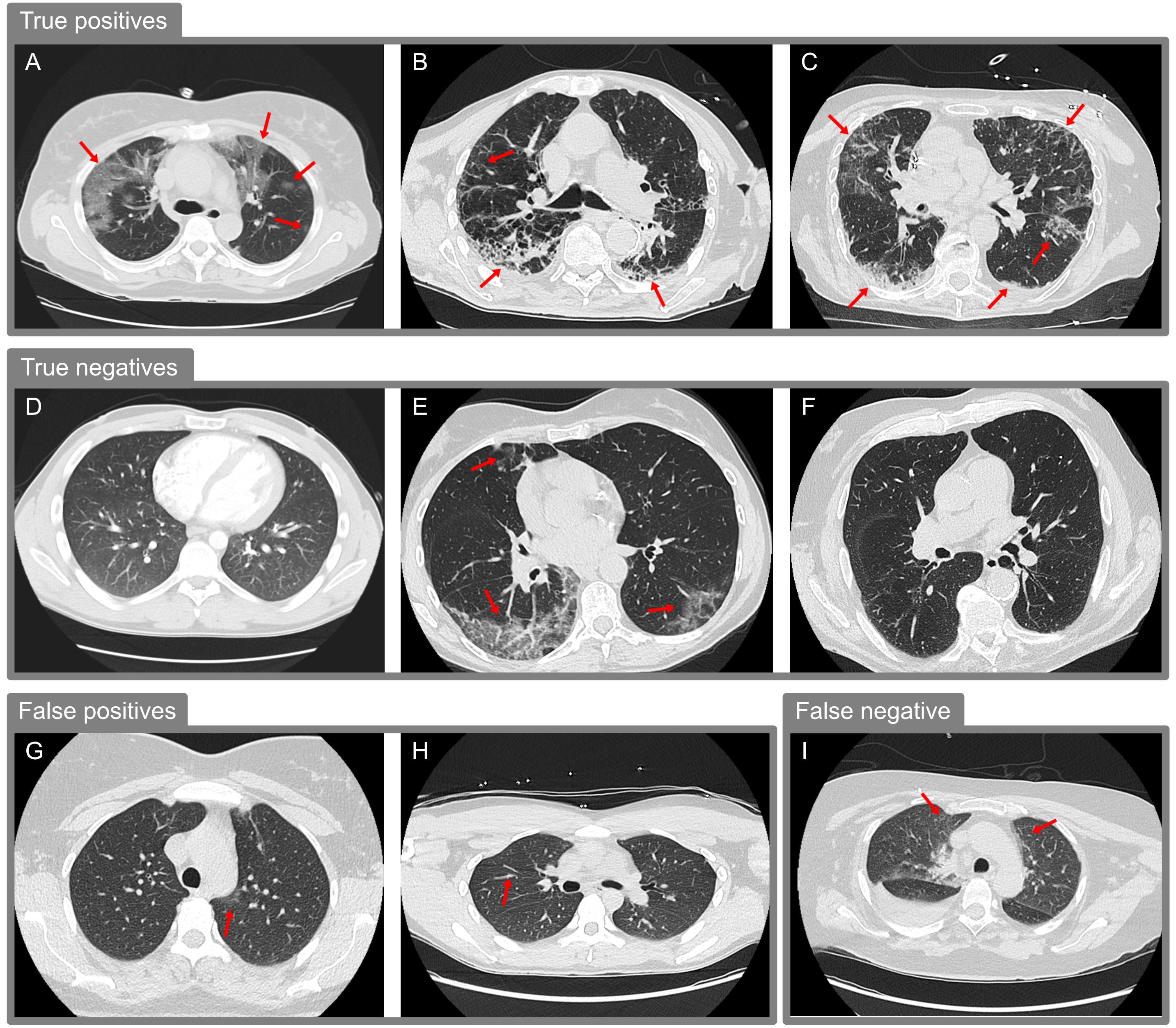}
		\caption{Examples of predictions. A-C: true positives; D-F: true negatives; G-H: false positives: I: false negative. Lesions are indicated with a red arrow.}
		\label{fig:examples_TP_TN_FP_FN}
	\end{figure}
\end{center}

Some examples of true predictions are shown in Figure~\ref{fig:examples_TP_TN_FP_FN}.(A-F) and of false predictions in Figure~\ref{fig:examples_TP_TN_FP_FN}.(G-I). Lesions are indicated with a red arrow. Though the patient in Figure~\ref{fig:examples_TP_TN_FP_FN}.E has large lesions, it is correctly classified as COVID-19 negative. For the case in Figure~\ref{fig:examples_TP_TN_FP_FN}.G, there is a small lesion in the left lung, the case  was wrongly classified as COVID-19 positive. For the false negative Figure~\ref{fig:examples_TP_TN_FP_FN}.I, the system missed the very subtle ground glass opacity.

\subsection{Computational complexity}
We analyzed the computational burden of our classification model along with the lung and lesions segmentation models. To carry out this analysis,  we first computed the amount of memory required by the models. Second, we calculated the number of floating point operations per second (FLOPs) executed in a forward pass of the models. Third, we leveraged the inference time to diagnose a patient, which includes the time required to obtain the lung and lesions segmentation in conjunction with the time required to classify and provide the explanations for the diagnose. In Table ~\ref{table:complexity} an overview is given of the number of parameters, FLOPS and inference time required by our method to diagnose COVID-19. As a whole, our proposal has 17.34 million parameters, requiring approximately 6.46 GB of memory. The models perform 5407.7 GigaFLOPS. Finally, the average inference time is 78 seconds, which was deemed acceptable, considering a radiologist spends approximately 15 minutes to diagnose a patient. Inference time was benchmarked using the NVIDIA TITAN Xp GPU, with $12$ GB of RAM.
\begin{table}[ht!]
	\caption{Computational complexity of the models employed to diagnose COVID-19 based on CT scans.}
	\label{table:complexity}
	\centering
	\resizebox{\columnwidth}{!}{\begin{tabular}{c c c c}
			\hline
			Method &Parameters (M) &FLOPs(G)&Inferece time (S) \\
			\hline
			Lung segmentation & 2.12& 4262.7 & 40 \\
			\hline
			Lesion segmentation &13.3 & 1113.8& 32 \\
			\hline
			Classification & 1.92 & 3.55 & 0.5 \\
			\hline
			Visual explanations & -& 27.65& 4.01 \\
			\hline
			Overall & 17.34 & 5407.7& 76.51 \\
			\hline
	\end{tabular}}
\end{table}


\section{Conclusions}
\label{sec:conclusions}
In this paper, we explore the CVAE as a generative model and found that it can achieve competitive results. In addition, the performance of the classification combined with the CVAE-based model was found to lead to comparatively accurate results in case of a smaller training set.  We further propose an extension of the basic CVAE model to incorporate side-information, in the form of a lesions segmentation mask, for learning disentangled representations. Combined with the explainable classification results, our system could be of interest for medical specialists to diagnose and understand better the COVID-19 infection at early stages via computed tomography. The developed tools will be integrated into the {\em icolung}\footnote{\url{ https://icometrix.com/services/icolung}} services, and can be used free of charge. The code for the developed models in this paper will be publicly available.

We did not perform hyper-parameter optimization for any of the experiments thus we expect the results to get better if further optimized. More tests on more difficult datasets, e.g. more classes and a higher variance, are desirable.

A limitation of the proposed CVAE method is that it utilizes binary labels for training and ignores the available hierarchical ground truth labels of the main patterns suggestive for COVID-19. An interesting extension of the method would be the use of the multi-class lesion segmentation and not the binarized form, as well as extending the CVAE to a hierarchical architecture to utilize the additional COVID-19 lesions category information for more accurate prediction.

%
%
%
%
%
%
%
%
%
%
%
%
%
%
%
%
%

\section*{Acknowledgement}
This research has been partially financed by the Flemish Government under the "Onderzoeksprogramma Artificiële Intelligentie (AI) Vlaanderen" programme (AI Research programme), by the FWO (project G0A4720N), and by the European Union under the Horizon 2020 Research and Innovation programme (project AI-based chest CT analysis enabling rapid COVID diagnosis and prognosis). We would like to thank MD.AI and Materialise to provide access to their platforms.

\bibliographystyle{model2-names}\biboptions{authoryear}
\bibliography{refs_v3}

\end{document}